\documentclass[pra,aps,epsfig,psfig,multicols,tightenlines,showpacs]{revtex4}
\usepackage{amsmath}
\usepackage{color}
\usepackage{amsfonts}
\usepackage{amsmath, amsthm, amssymb}
\usepackage{latexsym}
\usepackage{epsfig}
\usepackage{color}

\numberwithin{equation}{section}

\newcommand{\be}{\begin{equation}}
\newcommand{\ee}{\end{equation}}
\newcommand{\bea}{\begin{eqnarray}}
\newcommand{\eea}{\end{eqnarray}}
\newcommand{\beas}{\begin{eqnarray*}}
\newcommand{\eeas}{\end{eqnarray*}}
\newcommand{\ba}{\begin{array}}
\newcommand{\ea}{\end{array}}

\def\bge{\begin{eqnarray}}
\def\bgee{\begin{eqnarray*}}
\def\ege{\end{eqnarray}}
\def\egee{\end{eqnarray*}}

\begin{document}
\title{High-accuracy power series solutions with arbitrarily large radius of convergence for fractional nonlinear differential equations}
\author{U. Al Khawaja$^{1}$, M. Al-Refai$^2$, and Lincoln D. Carr$^3$}
\affiliation{\it $^1$Department of Physics, United Arab Emirates University, P.O. Box 15551, Al-Ain, United Arab Emirates\\
$^2$Department of Mathematical Sciences, United Arab Emirates University, P.O. Box 15551, Al-Ain, United Arab Emirates\\
$^3$Department of Physics, Colorado School of Mines, Golden, CO 80401, USA}
\date{\today}

\begin{abstract}
Fractional nonlinear differential equations present an interplay between two
common and important effective descriptions used to simplify high dimensional
or more complicated theories: nonlinearity and fractional derivatives.  These
effective descriptions thus appear commonly in physical and mathematical
modeling. We present a new series method providing systematic controlled
accuracy for solutions of fractional nonlinear differential equations.  The
method relies on spatially iterative use of power series expansions. Our
approach permits an arbitrarily large radius of convergence and thus solves the
typical divergence problem endemic to power series approaches. We apply our
method to the fractional nonlinear Schr\"odinger equation and its imaginary
time rotation, the fractional nonlinear diffusion equation. For the fractional
nonlinear Schr\"odinger equation we find fractional
generalizations of cnoidal
waves of Jacobi elliptic functions as well as a fractional bright soliton. For
the fractional nonlinear diffusion equation we find the combination of
fractional and nonlinear effects results in a more strongly localized solution
which nevertheless still exhibits power law tails, albeit at a much lower
density.
\end{abstract}


\maketitle

\section{Introduction}
Complexity is identified by a set of typical features~\cite{complexityNAS2009}, ranging from complex network descriptions~\cite{Newman2003} to chaos on strange attractors.  Fractional descriptions lay at the
foundation of many such features.  For instance, strange attractors typically have fractional
dimension.  Complex networks exhibit log-normal or power-law statistics also associated with
fractional derivatives~\cite{West2014}.  Fractional dimension is common in biological systems, e.g. the
folds of the brain, the space-filling curve of DNA, and the maximization of surface area for gas
exchange in the lungs and the branches of trees.  Although foundational descriptions of physical
phenomena typically take the form of partial differential equations (PDEs) or ordinary differential
equations (ODEs) with derivatives of integer order, for example the diffusion equation, the wave
equation, and the Schr\"odinger equation, it has been argued that effective PDE-based descriptions
of complex phenomena require \emph{fractional} partial derivatives~\cite{West2014,he1,mai,young}.
Indeed, the generalization of the diffusion equation to fractional derivatives has been found
highly effective in describing for instance the propagation of ground water through soil~\cite{benson2000}, the latter typically exhibiting multiscale properties.  To what extent are fractional
descriptions relevant to wave phenomena in general?

One natural context in which to examine this question is the \textit{nonlinear Schr\"odinger
equation} (NLSE).  The NLSE appears in many contexts including the experimentally and theoretically
well-established semiclassical effective limit of the quantum many-body description of
Bose-Einstein condensates~\cite{kevrekidisPG2008};  the slowly varying envelope approximation for
propagation of light in fiber optics~\cite{Agrawal2001}; and similar effective descriptions found for spin waves
in ferromagnetic films~\cite{kalinikos2000}.  In every case the NLSE is an \emph{effective} description, that is,
starting from a much more complete but difficult to solve theory, one obtains a single scalar
PDE/ODE that covers much of the observable phenomena in that system.  An outstanding physical
question is then the following.  On the one hand fractional differential equations provide an
effective description for complex multiscale phenomena in many contexts.  On the other hand
nonlinear PDEs/ODEs such as the NLSE are a highly useful simplification to reduce high dimensional
or otherwise difficult problems to effective nonlinear differential equations.  What is the
interplay between these two key classes of effective PDE/ODE descriptions of complex systems?

The fractional generalization of the NLSE in particular has several useful limits in which the
concept of effective theories can be carefully and rigorously explored.  First, the linear limit of
the Schr\"odinger equation is of course a well-understood problem.  Second, the fractional
generalization of the Schr\"odinger equation has been previously explored in many contexts~\cite{laskin2002,ZhangAndBellic2016}.  Third, the NLSE is an integrable equation with detailed solutions in terms
of solitons. Fourth, by performing an imaginary time rotation the Schr\"odinger equation can be
transformed to the diffusion equation.  Fifth, the fractional diffusion equation, perhaps even more
than the fractional Schr\"odinger equation, is a well-studied example with many special cases to draw
on. Thus the fractional NLSE provides an ideal mathematical context in which to explore the
interplay of nonlinearity and fractional derivatives in differential equations.

The basic concept of the fractional derivative is as old as calculus itself,
originating in a discussion between L'H\^{o}pital and Leibniz. In common usage
now there are several approaches to fractionalize the derivative and several
types of fractional derivatives have subsequently been introduced. The most
popular ones are the Riemann-Liouville and Caputo fractional derivatives.  The
two definitions differ only in the order of evaluation. In the Caputo
definition, we first compute an ordinary derivative then a fractional integral,
while in the Riemann-Liouville definition the operators are reversed. It is
often useful to apply the Caputo derivative in modeling, as mathematical
modeling of many physical problems requires initial and boundary conditions,
and these demands are satisfied using the Caputo fractional derivative.  As
fractional differential equations (FDEs) don't generally possess exact closed
form solutions, several analytical and numerical techniques have been
implemented to study these equations.  The series solution is one of the common
techniques in studying FDEs analytically and numerically.  For instance, the
variational iteration method (VIM)~\cite{Minc,obidat1}, the homotopy analysis
method (HAM)~\cite{Song,Sweilam}, and the Adomian decomposition method
(ADM)~\cite{Gejji,Jafari}, have been implemented to study several types of
FDEs.  For more details one can refer to~\cite{100} and references therein.
Recently, an efficient series solution was introduced for a class of nonlinear
fractional differential equations~\cite{alrefai1}. The idea is inspired by the
Taylor series expansion method but it overcomes the difficulty of computing
iterated fractional derivatives.

Despite these advances, there is no universal agreement on a series approach
that unifies solutions to fractional, nonlinear, and fractional nonlinear
differential equations.  In this paper, we provide such an approach.  Moreover,
previous power series methods suffer from divergence past a finite radius of
convergence.  For example, if one considers the power series expansion of the
bright soliton sech solution to the focusing NLSE, the radius of convergence is
$\pi/2$.  By extending the power series concept in a spatially iterative
approach, we are able to move the radius of convergence to arbitrarily large
values.  To see why this is so important physically, consider the central limit
theorem, which underpins statistical mechanics and indeed much of physical
measurement in terms of the normal distribution.  The central limit theorem
only holds for finite variance.  A well-known feature of the fractional
diffusion equation is the power law tails of its localized solutions, which
subsequently exhibit divergent variance.  Such ``fat-tailed'' distributions
lead to not-so-rare rare events and have practical outcomes such as the choice
of whether or not to store radioactive waste a given distance from a population
center~\cite{Benson2006}.  Our method allows us to benefit from the high
accuracy and analytical formulation of the power series approach while not
being subject to its typical problem with a physically limiting radius of
convergence. We are able to extend the radius of convergence arbitrarily far
into the tails of a localized solution, resulting in a clearer picture of the
kind of statistics we can expect from the hybridization of nonlinear and
fractional simplifications to physical problems.

\section{New power series ansatz}
\label{sec2} The common approach
to the power series problem for fractional differential equations
is an expansion of form
\begin{equation}
u(x)=\sum_{i=0}^{\infty}{a_i(x-x_0)^{i\alpha}} . \label{eqOld}\end{equation}
This approach works well for linear fractional equations.  However, for nonlinear
fractional equations Eq.~(\ref{eqOld}) has difficulties.
Instead, we propose the following fractional
power series as a solution:
\begin{equation}
u(x)=\sum_{i=0}^{n-1}{b_i(x-x_0)^i}+\sum_{i=0}^{\infty}{a_i(x-x_0)^{\alpha+i/q }}
\label{eq3},
\end{equation}
where $n=[\alpha]$ denotes the nearest integer larger than $\alpha$ and $x_0$ is the value about
which the expansion is performed. In comparison with the traditional expansion in powers of $i\alpha$ of Eq.~(\ref{eqOld}), our proposal in the second term of Eq.~(\ref{eq3}), $\alpha+i/q$, is more comprehensive in the sense that the former is a subset of the latter. For instance, with $\alpha=3/2$ and $q=2$, the term $x^{5/2}$ will not be present in a power series of $x^{i \alpha}$, while in our case this will be included. In addition, such terms couple to the integer power terms in the first sum of Eq.~(\ref{eq3}) and to the nonlinear term leading to nontrivial recursion relations. It is essential for the method developed here that we have the integer power terms in the first sum of Eq.~(\ref{eq3}), which renders the current proposed fractional power series to be a necessity  rather than a choice. Furthermore, when $q$ is even, our proposed power series will be valid also for the negative spatial domain.

In order to most effectively match the form of Eq.~(\ref{eq3}), we take the specific case of $\alpha=p/q$, where $p$ and
$q$ are positive integers. Adopting here the Caputo sense \cite{cap} for the fractional derivative,
defined by
\begin{equation}
\frac{d^\alpha}{d x^\alpha}=\left\{\begin{array}{cc}
\frac{1}{\Gamma(n-\alpha)}\int_{x_0}^x\frac{f^{(n)}(x')}{(x-x')^{\alpha-n}}\,dx',
& n-1<\alpha<n
\,,\\\\
\frac{d^n}{dx^n}f(x), &\alpha=n\,,
\end{array}
\right.
\label{cdef}
\end{equation}
where  $\Gamma(\cdot)$ is the well-known Gamma function, $f^{(n)}$ is the
$n^\mathrm{th}$ order derivative of $f$, $x>x_0$, and $x_0$ is a constant which in our case will
turn out to be the left side of the interval of spatial iteration in which we perform our power
series expansion.  For a power law, $f(x)=x^k$, where $k$ is an
integer, the Caputo fractional derivative reduces to
\begin{equation}
\frac{d^\alpha}{dx^\alpha}x^k=\left\{
\begin{array}{cc}
\frac{\Gamma(k+1)}{\Gamma(k+1-\alpha)}x^{k-\alpha},&k\ge\alpha\,,\\
\\
0,&k<\alpha\,.
\end{array}
\right.
\label{eq4}
\end{equation}
For a shifted
power law, $f(x)=(x-x_0)^k$, where $x_0$ is real, the above
expression is not applicable, as it would be in the case of
regular derivatives: the chain rule for fractional derivatives takes a different form than an in the integer case. To resolve this matter we resort to the definition in Eq.~(\ref{cdef}) to obtain
\begin{eqnarray}
&&\frac{d^\alpha}{d
x^\alpha}\left(x-x_0\right)^k=\\
&&\left\{\begin{array}{cc}
\frac{(k-1) k}{\Gamma (2-\alpha )}\left[ \frac{\left(-x_0\right){}^k
x^{1-\alpha } \left(-(\alpha -1) \, _2F_1\left(1,k-\alpha
;k;\frac{x_0}{x_0-x}\right)+k-1\right)}{(k-1) x_0
   (k-\alpha )}-\frac{\pi  \csc (\pi  \alpha ) \left(x-x_0\right){}^k \Gamma (k-1) \left(\frac{1}{x-x_0}\right){}^{\alpha }}{\Gamma (\alpha
   -1) \Gamma (k-\alpha +1)}\right], & 1<\alpha<2\,,\nonumber\\
   \\k(k-1)(x-x_0)^{k-2}, &\alpha=2\,,
\end{array}
\right. \label{cdef2}
\end{eqnarray}
where $_2F_1(\cdot)$ is the hypergeometric function and for simplicity we
have written out explicitly the solution of the integral to the case $1<\alpha\leq 2$; solutions for other values of $\alpha$ can be obtained in the same manner.  For physical applications $1<\alpha\leq 2$ is especially interesting as it interpolates between linear and quadratic disperson, i.e., between a phonon and a free particle.  Bogoliubov theory also accomplishes such an interpolation~\cite{fetterAL2001} but fractional derivatives are much more generic, e.g. in terms of capturing transport in a variety of multiscale media.  Also, the nonlinear Dirac equation, obtained for example by optical means or a BEC in a honeycomb lattice~\cite{carr2009g} near the K-point, has a first order spatial derivative and linear dispersion. It should be noted
here that Eq.~(\ref{cdef2}) is applicable for $k\ge\alpha$. For
$k<\alpha$, the fractional derivative equals zero for the case of
integer $k$. This is so, since for integer $k$, the expression
$(x-x_0)^k$ can be expanded in powers of $x$ that are all less
than $\alpha$, which, according to Eq.~(\ref{eq4}) will have zero
fractional derivative.

For our iterative method it will turn out to be useful to consider the special cases of $k=\alpha$ and
$k=\alpha+1$ with the result of the fractional derivative expanded around $x \to x_0^+$:
\begin{equation}
\frac{d^\alpha}{d
x^\alpha}\left(x-x_0\right)^\alpha=\left\{\begin{array}{cc}
c_0 +c_1 \log
\left(\frac{x-x_0}{x_0}\right)+\mathcal{O}(x-x_0),
& x_0>0\,,\\\\c_0, &x_0=0\,,
\end{array}
\right.\label{cdef3}
\end{equation}
\begin{eqnarray}
c_0 &=& -\frac{\pi  (\alpha -1) \alpha  \csc (\pi  \alpha )}{\Gamma (2-\alpha)}+c_1\left(-\frac{1}{\alpha -1}+\psi
^{(0)}(\alpha )+\gamma \right),\nonumber\\
c_1 &=& \frac{(-1)^{\alpha+1} (\alpha -1) \alpha}{\Gamma (2-\alpha )}\,,
\end{eqnarray}
where $\gamma$ is the Euler constant and $\psi ^{(0)}(\alpha )$ is
the digamma function defined by $\psi ^{(0)}(z)=d\,{\rm
Log}[\Gamma(z)]/dz$. For $x_0=0$ the $\alpha^{\rm th}$ derivative
given by Eq.~(\ref{cdef3}) is identical with that of Eq.~(\ref{eq4}) for $k=\alpha$, as it should be. For the case of
$k=\alpha+1$:
\begin{equation}
\frac{d^\alpha}{d
x^\alpha}\left(x-x_0\right)^{\alpha+1}=\left\{\begin{array}{cc}d_0+\left(d_1+d_2\log
   \left(\frac{x-x_0}{x_0}\right)\right)
   \left(x-x_0\right)+O[(x-x_0)]^2,
& x_0>0\,,\\\\d_3 x, &x_0=0\,,
\end{array}
\right.\label{cdef4}
\end{equation}
where
\begin{eqnarray}
d_0 &=& \frac{(-1)^{\alpha +1} \alpha
   (\alpha +1) x_0}{\Gamma (2-\alpha )},\nonumber\\
d_1&=&d_2
\left(\psi ^{(0)}(\alpha)+\gamma
-1\right)+d_3,\nonumber\\
d_2&=&\frac{\alpha \left(\alpha ^2-1\right)(-1)^{\alpha +1}}{\Gamma
(2-\alpha )},\nonumber\\
d_3 &=& d_2(-1)^{-(\alpha+1)}\pi  \csc (\pi \alpha
)\,.
\end{eqnarray}
We note that the expansion of Eq.~(\ref{cdef3}) diverges as $x\to x_0^+$ whereas Eq.~(\ref{cdef4}) approaches a constant.  This important observation
will be exploited in the construction of the power series that is
differentiable at $x_0$.

\section{Iterative solution method for the fractional nonlinear Schr\"odinger equation}
\label{sec:iterative}

We consider the following dimensionless fractional nonlinear Schr$\rm \ddot o$dinger equation
(FNLSE)
\begin{equation}
i\frac{\partial }{\partial t}U(x,t)=-\frac{1}{2}\,\frac{\partial^\alpha}{\partial x^{\alpha}}U(x,t)-\gamma\,|U(x,t)|^2 U(x,t)\,,
\label{eq1}
\end{equation}
where $\gamma$ is the strength of the nonlinearity.
The profile of the stationary solutions, $u(x)$, defined via $U(x,t)=u(x)\,\exp{(i\lambda\,t)}$, obeys the time-independent FNLDE
\begin{equation}
{1\over2}u-\frac{1}{2}\,\frac{d^\alpha }{dx^{\alpha}}u-\,u^3=0
\label{eq2},
\end{equation}
where we take $\lambda=1/2$ and $\gamma=1$ here and throughout such that for $\alpha=2$ the
fundamental soliton $u(x)={\rm sech{( {\it x} }) }$ is an exact solution of Eq.~(\ref{eq2}).  We
note that due to the nonlinear nature of Eq.~(\ref{eq2}) the factor of $\gamma$ can be absorbed
into the normalization without loss of generality, up to a sign: here we treat the focusing case.
Likewise, taking $\lambda=1/2$ incurs no loss of generality, as it is simply a choice of units of
time.

\begin{figure}
  \centering
  \includegraphics[width=15cm]{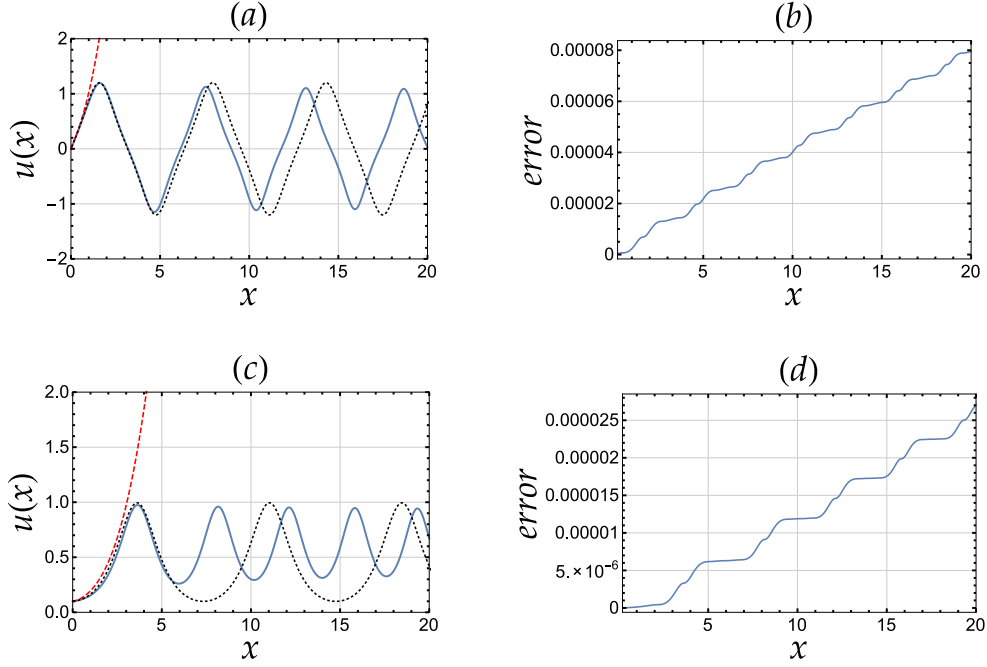}
  \caption{\textit{Iterative power series solutions of the fractional nonlinear
  Schr\"odinger equation.} (Left column) Fractional generalizations of the
  well-known cnoidal wave solutions of the NLSE for $\alpha=4/3$: (a) Jacobi
  elliptic cn-like with nodes; (c) dn-like without nodes.
  Converged iterative fractional nonlinear power series solution
  introduced for the first time in this paper (blue curves) is
  compared to the rapid divergence of the usual
  fractional power series (red dashed curves).  The well-known exact
  solutions for the NLSE are shown for comparison (black dotted curves).
  (Right column) (b), (d): Error incurred by our power series approach in
  each case increases arithmetically but remains below the predicted upper bound.
  The initial conditions are: $(u(0),u^\prime(0))=(0.0,0.8)$
  and $(0.1,0.01)$ for (a) and (c), respectively.
  The number of iterations used is $N=10^4$.}
  \label{fig1}
\end{figure}

As a result of Eq.~(\ref{eq4}), the $\alpha^{\rm th}$-derivative of the first
summation in Eq.~(\ref{eq3}) always vanishes. Therefore, the $b_i$ terms do not
couple to terms of lower powers and thus correspond to the initial conditions
such that for $0< \alpha\le1$ there is one initial condition, for $1<
\alpha\le2$ there are two initial conditions, etc.  Substituting the power
series solution, Eq.~(\ref{eq3}) with $x_0=0$, in a nonlinear equation such as
the FNLSE and using the Caputo fractional derivative, Eq.~(\ref{eq4}), it is
straightforward to derive the recursion relations for $a_i$ in terms of the
initial conditions $b_i$. However, the resulting power series solution suffers
from the typical problem of finite radius of convergence which depends on
$\alpha$ and equals $\pi/2$ for $\alpha=2$, as shown in Fig. \ref{fig1}.

To overcome this divergence problem, we implement the following iterative procedure.
First, we divide $x$ into $N$ segments, each of width $\Delta=x/N$.
Then we expand the solution starting with initial condition $x_0=0$, as given by Eq.~(\ref{eq3}). The resulting power series is then calculated at $x=\Delta$;
providing the initial conditions for the next expansion around $x_0=\Delta$, and so on.
However, two important issues need to
be addressed before obtaining a well-behaved and convergent iterative power
series solution using such an approach. The first is related to the logarithmic divergence
of the $\alpha^{\rm th}$ derivative of  $(x-x_0)^\alpha$, as shown
in Eqs.~(\ref{cdef3})-(\ref{cdef4}). The second is related to the convergence of the powers series
in the number of iterations, that is, how does the result depend on the number of grid points $N$ and choice of interval $\Delta$?

An immediate problem arises in this procedure: the $\alpha^{\rm th}$
derivative of the term $(x-x_0)^\alpha$ diverges logarithmically for
$x_0\ne0$, as shown in Eq.~(\ref{cdef3}). Fortunately, from Eq.~(\ref{cdef4}) we observe that the $\alpha^{\rm th}$
derivative of the term $(x-x_0)^{\alpha+1}$ does not suffer from such a divergence. The best approach is thus to expand $u(x)$ as follows:
\begin{equation}
u(x)=\left\{\begin{array}{ll}b_0+b_1x+a_1x^\alpha,
&x_0=0\,,\\
\\
b_0+b_1(x-x_0)+a_1(x-x_0)^{\alpha+1}, &x_0\ne0\,.
\end{array}
\right.\label{eq10}
\end{equation}
In this manner, one avoids the logarithmic divergence. Substituting
this expansion in Eq.~(\ref{eq2}), the recursion relation is
obtained as
\begin{equation}
a_1=\left\{\begin{array}{ll}\frac{\Gamma(1)}{\Gamma(\alpha+1)}(b_0-2b_0^3),
&x_0=0\,,\\
\\
\frac{\Gamma(2-\alpha)}{(-1)^{1+\alpha}\alpha(\alpha+1)x_0}(b_0-2b_0^3),
&x_0\ne0\,.
\end{array}
\right.\label{eq11}
\end{equation}
Then we calculate the series, Eq.~(\ref{eq10}) for $x_0=0$, at $x=\Delta$, from which we
calculate the {\it new} initial conditions
\begin{eqnarray}
b_0&\leftarrow&u(\Delta)=b_0+b_1\Delta+a_1\Delta^{\alpha}\,,\\
b_1&\leftarrow& u^\prime(\Delta)=b_1+\alpha a_1\Delta^{\alpha-1}
\label{eq9}.
\end{eqnarray}
The new values of $b_0$ and $b_1$ are then substituted back in Eq.~(\ref{eq10}), for $x_0=\Delta$, which in turn is then substituted in Eq.~(\ref{eq2}) to give the new recursion relation. It should be noted here that due to the homogeneity of the differential equation at hand, Eq.~(\ref{eq2}), the recursion relation $a_1(a_0)$ is essentially the same for all iterations, as given by Eq.~(\ref{eq10}) for $x_0\ne0$. This procedure is repeated $N$ times, leading to a power series free from the intrinsic divergence problem in the original series.  In the case of inhomogeneous differential equations, where the coefficients of $u$ and its derivatives are functions of $x$, the recursion relation $a_1(a_0)$ will be different for each iteration and has to be calculated at every iteration step.

However, the iterative procedure turns out to suffer from another serious problem. The solutions do not converge when $N\rightarrow\infty$. In a well-behaved iteration method, the solution should converge to a certain profile when $N$ is sufficiently increased. Careful investigation shows that this problem stems from the fact that the higher derivatives of the iterated power series are not continuous at the boundaries between the $i\Delta$ and $(i+1)\Delta$ expansions.  Basically, it is the iterative procedure itself that causes this problem. When we calculate the series $u(i\Delta)$ and its derivative $u^\prime(i\Delta)$ and pass these values to the $b_0$ and $b_1$ of the new series, we guarantee the continuity in the solution and its derivative only. The second derivative and higher derivatives are not continuous due to the fractional powers. This problem is absent in the case of integer power series but is an intrinsic problem in series with fractional powers. In the following we present a remedy to this difficulty.

Consider series zeroth-order series
\begin{equation}
u_0(x)=b_0+b_1x+a_1x^\alpha.
\end{equation}
Substituting $b_0=u_0(\Delta)$ and $b_1=u_0^\prime(\Delta)$, we obtain the first iterated series
\begin{equation}
u_1(x)=b_0+2b_1x+(\alpha+2)a_1x^\alpha.
\end{equation}
For the iterated series to be convergent with the number of iterations, $N$, it must be equal to the non-iterated series calculated at $N\Delta$, namely
\begin{equation}
u_N(\Delta)=u_0(N\Delta)\label{eq20}.
\end{equation}
Applying this condition for $N=1$,
\begin{equation}
b_0+b_1(2\Delta)+a_1(2\Delta)^\alpha=b_0+2b_1\Delta+(\alpha+2)a_1\Delta^\alpha.
\end{equation}
The zeroth- and first-order terms clearly satisfy this condition,
but the $\alpha^{\rm th}$-term acquires an additional factor of $(\alpha+2)$.
This requires the replacement $a_1\rightarrow a_1\times 2^\alpha/(\alpha+2)$
in the right hand side of this equation. Here we show the result of the next few iterations:
\begin{eqnarray}
u_2(x)&=&b_0+3b_1x+(3\alpha+3)a_1x^\alpha,\nonumber\\
u_3(x)&=&b_0+4b_1x+(6\alpha+4)a_1x^\alpha,\nonumber\\
u_4(x)&=&b_0+5b_1x+(10\alpha+5)a_1x^\alpha,\nonumber\\
u_5(x)&=&b_0+6b_1x+(15\alpha+6)a_1x^\alpha,\nonumber\\
\dots\nonumber\\
u_{N-1}(x)&=&b_0+N\,b_1x+[N+1+\frac{1}{2}N(N+1)\alpha]a_1x^\alpha,
\end{eqnarray}
where the closed form for the general term was found by inspection.
Applying the condition (\ref{eq20}) requires
\begin{equation}
a_1\rightarrow a_1\times \frac{N^\alpha}{N+1+\frac{1}{2}N(N+1)\alpha}\,.
\end{equation}
This condition is required to be satisfied in the limit of large $N$, which then takes the form
\begin{equation}
a_1\rightarrow a_1\times \frac{2}{\alpha}N^{\alpha-2}\label{eq30}\,.
\end{equation}
It should be noted that applying any of the last two conditions to
the series given by Eq.~(\ref{eq10}) for $x_0\ne0$ requires
replacing $\alpha$ by $\alpha+1$ in the last two equations. In
addition, Eqs. (\ref{eq10}) and (\ref{eq11}) show that $u(x)$
depends on $x_0$ also through $a_1\propto1/x_0$. Taking this into
account and repeating the iterative procedure described above but
now with $u(x)=b_0+b_1(x-x_0)+a_1(x-x_0)^{\alpha+1}$, which is the
second line of Eq.~(\ref{eq10}), we obtain finally the correction
replacement rule:
\begin{equation}
a_1\rightarrow a_1\times
(-1)^{\alpha+1}\Delta\frac{N^{\alpha+1}}{N+1+\frac{1}{2}N(N+1)(\alpha+1)}\,.
\end{equation}
The appearance of the factor $(-1)^{\alpha+1}$ suggests specific
fractional values of $\alpha=p/q$ with $q$ being an odd integer.
Otherwise, $a_1$ will become complex.

The resulting series for the FNLSE corresponds to a host of oscillatory and
localized solutions characterized by values of $b_0$ and $b_1$. The initial
slope of the solution, $b_1$, will not have a significant effect on the nature
of the solutions other than an overall shift along the $x$-axis. Inspection
shows that generally the solutions are oscillatory apart from localized
solutions obtained with some specific values of $b_0$.

\subsection{Fractional Cnoidal Waves}
In Fig.~\ref{fig1}, we plot two fundamental solutions of the FNLSE using this
iterative procedure. As clearly shown in this figure, the power series solution
without using the iterative procedure, (\ref{eq3}), diverges at $x\sim\pi/2$,
but after applying our iterative method the divergence is removed. Indeed,
implementing the correction procedure solves also the problem of lack of
convergence; the solutions are stable beyond a number of iterations greater
than about 2000. Another criterion that our solutions should satisfy is that
they must match the non-iterated series for a range less than the radius of
convergence, which is indeed the case, as Fig.~\ref{fig1} also shows. In Fig.
\ref{fig2}, we show a localized solution for different values of $\alpha$ with
$b_0$ tuned up to a certain precision in order to push oscillations away from
the origin to a certain distance. The analogous localized solution for the
$\alpha=2$ case is the fundamental soliton given by $u(x)={\rm sech}(x)$, which
is also plotted in Fig.~\ref{fig2}.

Error in the iterative series solution persists due to terminating the power
series at a finite number of terms, say $k$. It is instructive at this point to
estimate the magnitude of this error. Terminating the second summation in
Eq.~(\ref{eq3}) at $i=k$ results in an error of order
$\Delta^{\alpha+(k+1)/q}$. This error will be magnified $N$ times due to the
iterative procedure. Therefore, the estimated error in our method is of order
\begin{eqnarray}
{\rm error_{max}}\propto N\left(\frac{x}{N}\right)^{\alpha+(k+1)/q}
=N^{1-\alpha-(k+1)/q}\,x^{\alpha+(k+1)/q} \label{eq110}.
\end{eqnarray}
This estimate can be taken as an upper bound on the actual error, which we
define as
\begin{equation}
{\rm
error}=\frac{1}{x}\int_0^{x}\left|\frac{1}{2}u(x')-\frac{1}{2}
\frac{d^\alpha}{dx'^\alpha}u(x')-u(x')^3\right|^2\,dx'
\label{erreq},
\end{equation}
where $u(x')$ is the solution found by the iterative power series. In
the right column of Fig. (\ref{fig1}), we plot this error as a function of $x$.  We find the error increases linearly with $x$ but remains below the upper bound
(\ref{eq110}) for the parameters used.

\subsection{Fractional Solitons}
\begin{figure}
  \centering
  \includegraphics[width=15cm]{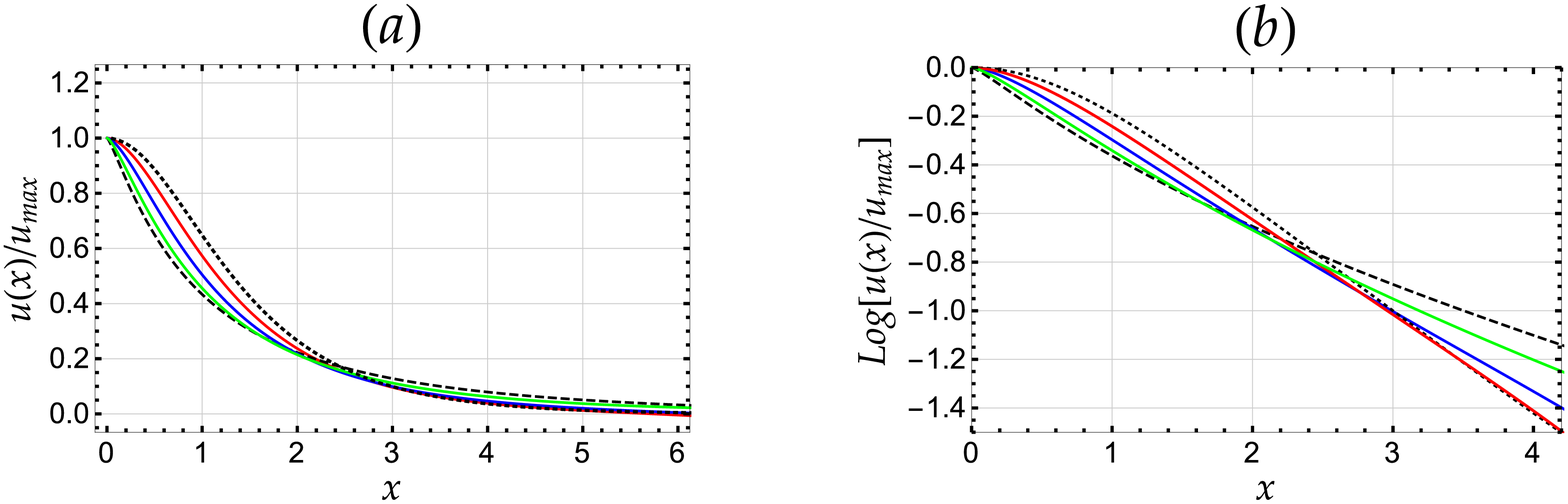}
  \caption{\textit{Fractional generalization of the bright soliton.} (a)
  Localized solution obtained via nonlinear fractional power series for values of $\alpha$ ranging between linear (dashed curve) and quadratic (dotted curve) dispersion: $\alpha=1.0001, 18/15, 22/15, 26/15, 2$. The usual $\alpha=2$ bright soliton, shown for comparison, is identical with ${\rm sech}(x)$.  Note that, for comparative purposes only, we renormalize the fractional soliton to its maximal value, plotting $u(x)/u_\mathrm{max}$ vs. $x$.  (b) The tails of such solutions remain exponentially decreasing and are therefore of finite variance.}
  \label{fig2}
\end{figure}

In Fig.~\ref{fig2}(a) we explore localized solutions more closely, where we
show their dependence on $\alpha$ as well as the structure of the tail.  We may
loosely call this solution the \textit{fractional soliton}, recognizing that
for now we have not shown stability.  We note that as $\alpha$ decreases the
soliton becomes narrower, i.e., is more strongly localized.  This can be
understood based on the balance between nonlinearity and dispersion that
supports a soliton.  Decreasing $\alpha$ effectively decreases the strength of
the dispersion, and as the nonlinearity is of focusing type, the soliton
becomes narrower.  However, we point out that this trend is opposite to that
predicted by the usual scaling argument based on a variational calculation of
the total energy.  The scaling behavior of Eq.~(\ref{eq2}) obtained in such a
manner would naively be $f\propto 1/\ell^\alpha - 1/\ell$, with $\ell$ an
estimate of the width of the fractional soliton, as follows also from the
$1/\sqrt{\ell}$ units of $u(x)$.  However, solving for $df/d\ell=0$ results in
$\ell_\mathrm{min} = \alpha^{1/(\alpha-1)}$, predicting a decreasing width for
increasing $\alpha$, clearly in contradiction to the more accurate power series
approach.  In the future it would be useful to develop a proper variational
analysis via a Lagrangian variational dynamical approach~\cite{perez1997} to
obtain a more precise estimate of fractional soliton widths and breathing
frequencies.  This requires rethinking the energy contribution of the
fractional derivative.

As illustrated in Fig.~\ref{fig2}(b) our iterative power series approach finds
that the tail is exponential in nature, in contrast to the well-known power-law
behavior for the fractional diffusion equation discussed in
Sec.~\ref{sec:diffusion}.  Although it would seem the tail of a localized
solution of any nonlinear fractional equation should approach the tail of the
corresponding linear fractional equation, this is not at all obvious since a
tail grows rapidly flat, and one must closely examine the balance between
shrinking nonlinearity and decreasing derivative on the tail.  We find the
prefactor in the exponential decreases monotonically with $\alpha$.

\section{Application to the fractional nonlinear diffusion equation}
\label{sec:diffusion}

It is well known that the imaginary time rotation of the Schr\"odinger equation $\tau = - i t$ produces the diffusion equation.  Both nonlinear and fractional extensions of these equations are connected in the same way, and thus the case study for our power series in Sec.~\ref{sec:iterative} is easily applied to the fractional nonlinear diffusion equation.  This is an especially interesting case because the fractional diffusion equation has well-known power law tails exhibiting a divergent variance.

The fractional diffusion equation takes the form
\begin{equation}
\frac{\partial u(x,\tau)}{\partial \tau}=\frac{\partial^\alpha
u(x,\tau)}{\partial x^\alpha} \label{eq200} \,.\end{equation}
This
well-studied equation is usually solved by the Fourier transform
method.  For $\alpha=2$ the solution is the Gaussian function
\begin{equation}
u(x,\tau)=\frac{c}{\sqrt{\tau}}e^{-\frac{x^2}{4\tau}}\,,
\end{equation}
where $c$ is an arbitrary real
constant. For the fractional case, the solution is still localized, but
with an algebraically
decaying tail of form $|x|^{-(\alpha+1)}$ \cite{benson2000}. This is a distinctive feature of the fractional
diffusion equation which we aim at obtaining using the present
method.

\begin{figure}
  \centering
  \includegraphics[width=15cm]{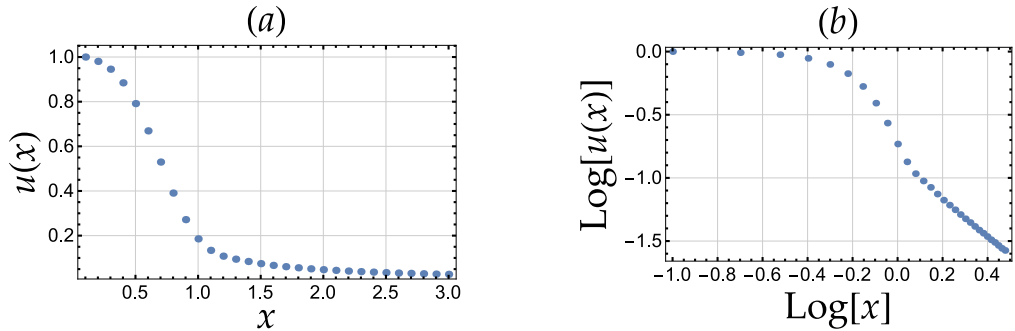}
  \caption{\textit{Power law tails in fractional diffusion equation.}  Solution of the fractional diffusion equation, Eq.~(\ref{eq200}) at $\tau=1$ and for $\alpha=4/3$.
  We used the initial conditions $u(0)=1$ and $u^\prime(0)=0$. Number of iterations used is $N=30$.}
  \label{fig3}
\end{figure}

Based on the special case of $\alpha=2$, a simple scaling argument may be used to deduce that the solution
scales as
\begin{equation}
u(x,\tau)=\frac{1}{\sqrt{\tau}}\,v\left(\frac{x}{2\sqrt{\tau}}\right)\,,
\end{equation}
which suggests the following series expansion:
\begin{equation}
u(x,\tau)=\frac{1}{\sqrt{\tau}}\left[b_0+b_1\left(\frac{x}{2\sqrt{\tau}}\right)
+a_1\left(\frac{x}{2\sqrt{\tau}}\right)^\alpha\right] \label{eq300}.
\end{equation}
Applying the iterative power series method as described above, we obtain the
desired results, as shown in Fig. \ref{fig3}. On a log-log scale, Fig.
\ref{fig3}(b) shows clearly that the tail is decaying algebraically, as
expected, with the correct exponent of the power law. Due to the fact that the
diffusion equation is linear a rather small number of iterations, $N=30$, is
enough to obtain solutions that converge against increasing the number of
iterations. While the algebraic decay of the tail is accounted for qualitatively, a quantitative account requires a more rigorous treatment in which we do not use the above-mentioned specific scaling. This in turn requires a generalisation of the present method to partial differential equations, which we leave to future work.

\begin{figure}
  \centering
  \includegraphics[width=10cm]{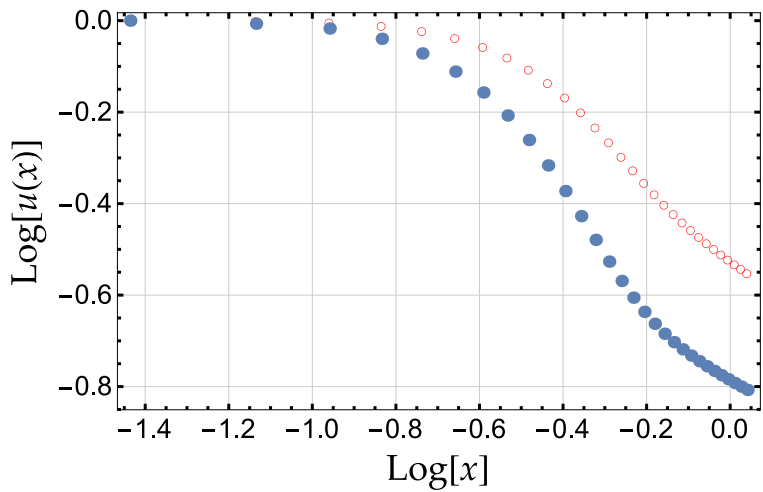}
  \caption{\textit{Localized solutions in the fractional nonlinear diffusion equation.} Solution of the nonlinear fractional diffusion
  equation, Eq.~(\ref{eq300}) at $\tau=1$, for $\alpha=4/3$, and $\gamma=1$ (Blue filled
  circles). Red empty circles correspond to the linear case, $\gamma=0$.
  We used the initial conditions $u(0)=1$ and $u^\prime(0)=0$. Number of iterations used is $N=2000$.}
  \label{fig4}
\end{figure}

Finally, we apply our method to the nonlinear
fractional diffusion equation
\begin{equation}
\frac{\partial u(x,\tau)}{\partial \tau}=\frac{\partial^\alpha
u(x,\tau)}{\partial x^\alpha}-\gamma u(x,\tau)^3
\label{eq300},\end{equation} where $\gamma$ is a positive constant.
The effect of the cubic nonlinear term is shown in Fig. \ref{fig4}
where we compare with the linear case. The algebraic decay of the
tail persists in the nonlinear case but at a much lower density than
for the linear case.  Thus the focusing nonlinearity localizes the solution much more strongly than in the linear case, but in the end the power law is the same, indicating that the fractional derivative dominates over the nonlinearity asymptotically just like in the integer derivative case.

\section{Conclusions}

We developed a new iterative power series approach to solve fractional
nonlinear differential equations.  The use of iteration allows one to overcome
the usual radius of convergence problems associated with power series.  We
provided an explicit upper bound on the error of our method in the total number
of spatial iterations and the grid size, or spatial interval.  We applied our
method to the fractional nonlinear Schr\"odinger equation (FNLSE) and its
imaginary time rotation, the fractional nonlinear diffusion equation.  We found
fractional generalizations of cnoidal waves and introduced the fractional
soliton for the focusing FNLSE.  For the fractional nonlinear diffusion
equation we showed power law tails persist, although at a much lower density
than in the linear case: thus fractional dispersion dominates nonlinearity
asymptotically.  In future work the method can be applied to a wide variety of
nonlinear fractional equations, the most obvious being the defocusing FNLSE
case where one can study the properties of fractional dark solitons.

\section*{Acknowledgements}
UAK acknowledges the support of UAE University through the grants UPAR(6) and UPAR(7) and fruitful discussions during his stay at the Colorado School of Mines.  LDC acknowledges support of the US National Science Foundation and Air Force Office of Scientific Research and fruitful discussions during his stay at UAE University.

\end{document}